\documentstyle[preprint,revtex]{aps}
\def\beq{\begin{eqnarray}}
\def\eed{\end{eqnarray}}
\begin{document}
\draft

\begin{title}
\begin{center}
{Charge Dynamics from Copper Oxide Materials}
\end{center}
\end{title}
\author{Shiping Feng$^{1,2,3}$, Zhongbing Huang$^{2}$}
\begin{instit}
$^{1}$CCAST (World Laboratory) P. O. Box 8730, Beijing 100080, China and \\
$^{2*}$Department of Physics, Beijing Normal University, Beijing
 100875, China and \\
$^{3}$National Laboratory of Superconductivity, Academia Sinica,
 Beijing 100080, China \\
\end{instit}
\begin{abstract}
The charge dynamics of the copper oxide materials
in the underdoped and optimal doped regimes is studied within the
framework of the fermion-spin theory. The conductivity spectrum shows
the non-Drude behavior at low energies and unusual midinfrared peak, and
the resistivity exhibits a linear behavior in the temperature, which are
consistent with experiments and numerical simulations.
\end{abstract}

Since the discovery of the copper oxide superconductors, a significant
body of reliable and reproducible data has been accumulated by using
many probes \cite{n1}, which show that the most remarkable expression
of nonconventional physics of copper oxide materials is found in the
normal-state. The normal-state properties exhibit a number of anomalous
properties in sense that they do not fit in the conventional Fermi-liquid
theory, and some properties mainly depend on the extend of dopings
\cite{n1}. The undoped copper oxide materials are antiferromagnetic Mott
insulators, and the antiferromagnetic long-range-order (AFLRO) disappears
when the hole doping concentration exceeds some critical value
(about $5\%$) \cite{n2}. The charge response in the underdoped and optimal
doped regimes, as manifested by the optical conductivity and resistivity,
has been extensively studied experimentally in the copper oxide materials,
as well as theoretically within some strongly correlated models
\cite{n3,n4,n5,n6,n7}. From experiments testing normal-state properties,
the optical conductivity shows the non-Drude behavior at low energies and
anomalous midinfrared band in the charge-transfer gap, the resistivity
exhibits a linear behavior in the temperature in the optimal doped regime
and a nearly temperature linear dependence with deviations at low
temperatures in the underdoped regime \cite{n1,n3,n4}.  These optical
properties indicate that the strong correlations are very important to
the electronic structure.

Among the microscopic models the most helpful for the discussion of
the normal-state properties in the copper oxide materials is the $t$-$J$
model acting on the space with no doubly occupied sites \cite{n8}.
In order to account for real experiments under the $t$-$J$ model, the
crucial requirement is to impose the electron local constraint \cite{n9}.
Recently a fermion-spin theory based on the charge-spin separation is
proposed \cite{n10} to incorporated this constraint. In this approach, the
electron on-site local constraint for single occupancy is satisfied even in
the mean-field approximation (MFA). Within the fermion-spin theory, it has
been shown \cite{n11} that AFLRO vanishes around doping $\delta =5\%$ for
the reasonable value of the parameter $t/J=5$, and the mean-field theory
\cite{n12} in the underdoped and optimal doped regimes without AFLRO has
been also developed, which has been applied to study the photoemission
and electron dispersion. In this paper, we try to study the charge
dynamics of the copper oxide materials within this fermion-spin theory.
We follow the fermion-spin formulism \cite{n10,n12} and decompose the
electron operators as $C_{i\uparrow}=h^{\dagger}_{i}S^{-}_{i}$ and
$C_{i\downarrow}=h^{\dagger}_{i}S^{+}_{i}$, where the spinless fermion
operator $h_{i}$ keeps track of the charge (holon), while the pseudospin
operator $S_{i}$ keeps track of the spin (spinon). The mean-field order
parameters are defined \cite{n12} as
$\chi=\langle S_{i}^{+}S_{i+\eta }^{-}\rangle =\langle S_{i}^{-}
S_{i+\eta}^{+}\rangle$, $\chi_{z}=\langle S_{i}^{z}S_{i+\eta }^{z}\rangle$,
$C=(1/Z^{2})\sum_{\eta ,\eta ^{\prime }}\langle S_{i+\eta }^{+}
S_{i+\eta ^{\prime}}^{-} \rangle$, $C_{z}=(1/Z^{2})
\sum_{\eta ,\eta ^{\prime }}\langle S_{i+\eta }^{z}S_{i+\eta ^{\prime }}^{z}
\rangle$, and $\phi =\langle h_{i}^{\dagger}h_{i+\eta }\rangle$ with
$\hat{\eta }=\pm \hat{x},\pm \hat{y}$, and $Z$ is the number of nearest
neighbor sites. In this case, the low-energy behavior can be described
\cite{n12} by the effective Hamiltonian $H=H_t+H_J$ with
\begin{mathletters}
\begin{eqnarray}
H_t=-t \sum_{i\eta}h_{i}h_{i+\eta}^{\dagger}(S_{i}^{+}S_{i+\eta}^{-}+
S_{i}^{-}S_{i+\eta}^{+})+h.c.+\mu \sum_{i}h_{i}^{\dagger}h_{i} , \\
H_J=J_{eff}\sum_{i\eta}[{1 \over 2}(S_{i}^{+}
S_{i+\eta}^{-}+S_{i}^{-}S_{i+\eta}^{+})+S_{i}^{z}S_{i+\eta}^{z}],~~~~~~~~
\end{eqnarray}
\end{mathletters}
where $J_{eff}=J[(1-\delta )^2-\phi ^2]$, and $\mu$ is the chemical
potential which enforce $\langle h_{i}^{\dagger}h_{i}\rangle =\delta$.

The spinon and holon may be separated at mean-field level, but they are
strongly coupled beyond MFA due to fluctuations \cite{n13}. For discussing
the optical conductivity and resistivity, we now need to consider the
dynamical effects beyond MFA. Since an electron is represented by the
product of a holon and a spinon in the fermion-spin theory \cite{n10,n12},
the external field can only be coupled to one of them. According to
Ioffe-Larkin combination rule \cite{n14}, we can separately calculate the
contributions to the conductivity from holons $\sigma_{h}(\omega)$ and
spinons $\sigma_{s}(\omega)$, and the total conductivity is given by
$\sigma^{-1}(\omega)=\sigma_{h}^{-1}(\omega)+\sigma_{s}^{-1}(\omega)$.
In the present theoretical framework, the current densities of spinons and
holons are defined as $j_{s}=te\phi\sum_{i\eta }\hat{\eta}(S_{i}^{+}
S_{i+\eta}^{-}+S_{i}^{-}S_{i+\eta}^{+})$ and $j_{h}=2te\chi\sum_{i\eta}
\hat{\eta}h_{i+\eta }^{\dagger}h_{i}$, respectively. A formal calculation
for the spinon part shows that there is no the direct contribution to the
current-current correlation from spinons, but the strongly correlation
between holons and spinons is considered through the spinon's order
parameters $\chi$ , $\chi_{z}$ , $C$, and $C_{z}$ entering in the holon
part of the contribution to the current-current correlation, which
means that the holon moves in the background of spinons, and the cloud of
distorted spinon background is to follow holons, therefore the dressing of
the holon by spinon excitations is the key ingredient in the explanation of
the transport property. In this case, the conductivity is expressed as
$\sigma(\omega)=-{\rm Im}\Pi_{h}(\omega)/\omega$ with the holon-holon
correlation function $\Pi_{h}(i\omega_{n})=-(2te\chi Z)^2(1/N)
\sum_{k}\gamma_{sk}^2(1/\beta)\sum_{i\omega_{m'}}
g(k,i\omega_{m'}+i\omega_{n})g(k,i\omega_{m'})$, where
$\gamma_{sk}=(1/2)({\rm sin}k_{x}+{\rm sin}k_{y})$ and the
holon Green's function is defined as $g_{ij}(\tau -\tau ^{\prime })=
-\langle T_{\tau}h_{i}(\tau)h_{j}^{\dagger}(\tau ^{\prime})\rangle$. In
this paper, we limit the spinon part to the first-order (mean-field level)
since some physical properties can be well described at this level \cite{n12},
and there is no direct contribution to the charge dynamics from spinons as
mentioned above. However, the second-order correction for the holon is
necessary for a proper description of holon motion due to antiferromagnetic
fluctuations. The mean-field spinon Green's functions $D^{(0)}(k,\omega)$
and $D^{(0)}_{z}(k,\omega)$ and mean-field holon Green's function
$g^{(0)}(k,\omega)$ have been given in Ref. \cite{n12}. The second-order
holon self-energy diagram from the spinon pair bubble is shown in Fig. 1.
Since the spinon operators obey the Pauli algebra, we map the spinon
operator into the CP$^{1}$ fermion representation or spinless-fermion
representation in terms of the 2D Jordan-Wigner
transformation \cite{n15} for the formal many particle perturbation
expansion. After then the spinon Green's function in the holon self-energy
diagram shown in Fig. 1 is replaced by the mean-field spin Green's
function $D^{(0)}(k,\omega)$, and therefore the second-order holon
self-energy is evaluated as
\begin{eqnarray}
\Sigma_{h}^{(2)}(k,\omega)=(Zt)^{2}{1\over N^2}\sum_{pp'}(\gamma_{p'-k}+
\gamma_{p'+p+k})^{2}B_{p'}B_{p+p'}\times \nonumber \\
\left ( 2{n_{F}(\xi_{p+k})
[n_{B}(\omega_{p'})-n_{B}(\omega_{p+p'})]-n_{B}(\omega_{p+p'})
n_{B}(-\omega_{p'})\over \omega +\omega_{p+p'}-\omega_{p'}-
\xi_{p+k}+i0^{+}} \right. \nonumber \\
+{n_{F}(\xi_{p+k})[n_{B}(\omega_{p+p'})-n_{B}(-\omega_{p'})]+
n_{B}(\omega_{p'})n_{B}(\omega_{p+p'})\over \omega +\omega_{p'}+
\omega_{p+p'}-\xi_{p+k}+i0^{+}} \nonumber \\
\left. -{n_{F}(\xi_{p+k)}[n_{B}(\omega_{p+p'})
-n_{B}(-\omega_{p'})]-n_{B}(-\omega_{p'})n_{B}(-\omega_{p+p'}) \over
\omega -\omega_{p+p'}-\omega_{p'}-\xi_{p+k}+i0^{+}}\right ),
\end{eqnarray}
where $\gamma_{k}=(1/Z)\sum_{\eta}e^{ik\cdot \hat{\eta}}$,
$\epsilon =1+2t\phi/J_{eff}$, $B_k=ZJ_{eff}[(2\epsilon \chi_z+\chi)
\gamma_{k}-(\epsilon \chi +2\chi_z)]/\omega_{k}$,
$n_{F}(\xi_{k})$ and $n_{B}(\omega_{k})$ are the Fermi and Bose
distribution functions, respectively, the mean-field holon excitation
spectrum $\xi_{k}=2Z\chi t\gamma_{k}+\mu$, and the mean-field
spinon excitation spectrum $\omega_{k}$ is given in Ref. \cite{n12}.
Then the full holon Green's function is obtained as
$g^{-1}(k,\omega)=g^{(0)-1}(k,\omega)-\Sigma_{h}^{(2)}(k,\omega)$.
We emphasize that the local constraint of the $t$-$J$ model has been
treated exactly in the mean-field theory \cite{n12}, and it is still
satisfied in the above perturbation expansion based on this mean-field
theory.

With the help of the full holon Green's function, the optical conductivity
can be obtained as
\begin{eqnarray}
\sigma (\omega )={1\over 2}(2te\chi Z)^2{1\over N}\sum_k\gamma_{sk}^{2}
\int^{\infty}_{-\infty}{d\omega'\over 2\pi}A_{h}(k,\omega'+\omega)
A_{h}(k,\omega'){n_{F}(\omega'+\omega)-n_{F}(\omega') \over \omega},~~~~
\end{eqnarray}
with the holon spectral function $A_{h}(k,\omega)=-2{\rm Im}g(k,\omega)$.
We have performed a numerical calculation for this optical conductivity
(3), and the results at the doping $\delta = 0.06$ (solid line) and
$\delta = 0.12$ (dashed line) for the parameter $t/J=2.5$ with the
temperature $T=0$ are shown in Fig. 2, where the charge $e$ has been set
as the unit. For comparison, the result from numerical simulation
based on the Lanczos
diagonalization \cite{n5} of a $4\times 4$ cluster with a single doped
hole ($\delta \approx 0.06$) for $t/J=2.0$ is also shown in Fig .2
(dot-dashed line). Although the infrared properties of the copper
oxide materials are very complicated, some qualitative features such as
away from half-filling, weight appears inside the charge-transfer gap of
the undoped materials, defining the midinfrared band and the conductivity
decays as $\rightarrow 1/ \omega$ at low energies, seem common to all
copper oxide materials \cite{n1,n3,n4}. The present theoretical result
shows that there are a low-energy peak at $\omega <0.5t$ separated by
a gap or pseudogap $\approx 0.5t$ from the broad absorption band
(midinfrared band) in the conductivity spectrum. The analysis of the
result indicates that the conductivity decays is $\rightarrow 1/\omega$
(non-Drude falloff) at low energies ($\omega <0.5t$), the midinfrared
peak is doping dependent and the peak is shifted to lower energy with
increased doping, which is in qualitative agreement with the experiment
\cite{n4} and numerical simulation \cite{n5,n6}.

It has been shown that the Drude weight can be used as an order parameter
for metal-insulator transitions \cite{n16}, which converges exponentially
to zero for the insulator, and nonzero constant for the metal. Following
the standard discussions \cite{n17}, the Drude weight $D$ is obtained in the
present case as
\begin{eqnarray}
D_{n}={\langle -T\rangle\over 4} +\lim_{i\omega_{n}\rightarrow 0}
(2\chi t)^{2}{1\over N}\sum_{k}{\rm sin}^{2}k_{x}\int^{\infty}_{-\infty}
{d\omega\over 2\pi}{d\omega'\over 2\pi}A_{h}(k,\omega)A_{h}(k,\omega')
{n_{F}(\omega')-n_{F}(\omega)\over i\omega_{n}+\omega'-\omega},~~~~
\end{eqnarray}
where $D_{n}=D/2\pi e^{2}$, and the total kinetic energy per site
$\langle T\rangle =8t\chi \phi$. The numerical results of eq. (4) for
$t/J=2.5$ (solid line) is plotted in Fig .3. The result shows that the
Drude weight grows rapidly with increasing doping $\delta$, and the
doping dependence is nearly linear in the underdoped and optimal doped
regimes, which is qualitative consistent with the exact diagonalization
result \cite{n18}.

The resistivity is expressed as $\rho=1/\sigma_{dc}$, where the dc
conductivity $\sigma_{dc}$ is obtained from Eq. (3) as
$\sigma _{dc}=\lim_{\omega \rightarrow 0}\sigma (\omega)$.
This resistivity has been evaluated numerically and
the results in the doping $\delta =0.06$ and $\delta =0.10$ for $t/J=2.5$
are plotted in Fig. 4(a) and Fig. 4(b), respectively, which show
that in the underdoping, the resistivity indeed exhibits a nearly
temperature linear dependence with deviations at low temperature, and a
very good linear behavior in the optimal doping, which also is in
agreement with the experiments \cite{n1,n3,n4}.

In the present theory, the basic low-energy excitations are holons and
spinons. However, our results show that the unusual non-Drude type optical
behavior at low energies and midinfrared band, and linear temperature
dependence of the resistivity are mainly caused by the holons in the copper
oxide sheets, which are strongly renormalized because of the interactions
with fluctuations of the surrounding spinon excitations. The $1/\omega$
decay of the conductivity at low energies is closely related with the
linear temperature resistivity, this reflects an anomalous frequency
dependent scattering rate proportional to $\omega$ instead $\omega^{2}$
as would be expected for the conventional Fermi-liquid theory, which is
consistent with the Luttinger-liquid theory \cite{n19} and phenomenological
marginal Fermi-liquid theory \cite{n20} Our study seems to confirm that
the strongly correlated $t$-$J$ model is possibly sufficient for the
understanding of the normal-state properties of the copper oxide materials.

In summary, we have studied the charge dynamics of the copper oxide
materials in the underdoped and optimal doped regimes within the $t$-$J$
model. It is shown that the strongly correlated renormalization effects of
the holon motion due to spinon fluctuations are very important for the
charge dynamics. The optical conductivity, Drude weight and
resistivity are discussed, and the results are qualitative consistent with
the experiments and numerical simulations.

\acknowledgments
This work is supported by the National Science Foundation
Grant No. 19474007 and the Trans-Century Training Programme Foundation
for the Talents by the State Education Commission of China.

\references{

\bibitem [*] {add} Mailing address.

\bibitem{n1} See, e. g., B. Batlogg, {\it in High Temperature
Superconductivity}, Proc. Los Alamos Symp., 1989, K. S. Bedell
{\it et al}., eds. (Addison-Wesley, Redwood City, California, 1990);
B. Batlogg {\it et al.}, Physica C {\bf 235-240}, 130 (1994), and
references therein.

\bibitem{n2} A. P. Kampf, Phys. Rep. {\bf 249}, 219 (1994), and
references therein.

\bibitem{n3} H. Takagi {\it et al}., Phys. Rev. Lett. {\bf 69},
2975 (1992); T. Ito, K. Takenaka, and S. Uchida, Phys. Rev. Lett.
{\bf 70}, 3995 (1993); W. Si and Z. X. Zhao (private communication).

\bibitem{n4} S. Uchida, Mod. Phys. Lett. B{\bf 4}, 513 (1990);
T. Timusk {\it et al}., Physica C{\bf 162}-{\bf 164}, 841 (1989).

\bibitem{n5} W. Stephan and P. Horsch, Phys. Rev. B{\bf 42},
8736 (1990).

\bibitem{n6} A. Moreo and E. Dogotto, Phys. Rev. B{\bf 42}, 4786
(1990).

\bibitem{n7} E. Dagotto, Rev. Mod. Phys. {\bf 66}, 763 (1994), and
references therein.

\bibitem{n8} P. W. Anderson, Science {\bf 235}, 1196 (1987);
F. C. Zhang and T. M. Rice, Phys. Rev. B{\bf 37}, 3759 (1988).

\bibitem{n9} L. Zhang, J. K. Jain, and V. J. Emery, Phys. Rev. B{\bf 47},
3368 (1993); Shiping Feng {\it et al.}, Phys. Rev. B{\bf 47}, 15192
(1993).

\bibitem{n10} Shiping Feng, Z. B. Su, and L. Yu, Phys. Rev. B
{\bf 49}, 2368 (1994); Mod. Phys. Lett. B{\bf 7}, 1013 (1993).

\bibitem{n11} Shiping Feng, Yun Song, and Zhongbing Huang,
Mod. Phys. Lett. B{\bf 10}, 1301 (1996).

\bibitem{n12} Shiping Feng and Yun Song, Phys. Rev. B {\bf 55}, 642
(1997).

\bibitem{n13} See, e.g., L. Yu, in {\it Recent Progress in Many-Body
Theories}, edited by T. L. Ainsworth {\it et al}. (Plenum, New York,
1992), Vol. 3, p. 157.

\bibitem{n14} L. B. Ioffe and A. I. Larkin, Phys. Rev. B {\bf 39}, 8988
(1989).

\bibitem{n15} E. Mele, Phys. Scr. T {\bf 27}, 82 (1988); E. Fradkin,
Phys. Rev. Lett. {\bf 63}, 322 (1989).

\bibitem{n16} W. Kohn, Phys. Rev. {\bf 133}, A171 (1964).

\bibitem{n17} D. J. Scalapino, S. R. White, and S. C. Zhang, Phys. Rev.
Lett. {\bf 68}, 2830 (1992).

\bibitem{n18} E. Dagotto {\it et al.}, Phys. Rev. B {\bf 45}, 10741
(1992).

\bibitem{n19} P. W. Anderson, Phys. Rep. {\bf 184}, 195 (1989).

\bibitem{n20} C. M. Varma {\it et al}., Phys. Rev. Lett. {\bf 63},
1996 (1989).
}

\figure{The holon's second-order self-energy diagram. The solid and
dashed lines correspond to the holon and spinon propagators,
respectively.}

\figure{The optical conductivity at the doping $\delta=0.06$ (solid line)
and $\delta=0.12$ (dashed line) for $t/J=2.5$ with the temperature $T=0$.
The dot-dashed line is the result from the numerical simulation on a
$4\times 4$ cluster with a single doped hole for $t/J=2.0$ \protect\cite{n5}.}

\figure{Drude weight, in the unit of $t$, as a function of doping
$\delta$ for $t/J=2.5$.}

\figure{The electron resistivity at the parameter $t/J=2.5$ for (a) the
doping $\delta=0.06$ and (b) $\delta=0.10$. The small wiggles are the
finite-size effect.}

\end{document}